\documentclass[sn-mathphys]{sn-jnl}

\jyear{2021}%

\theoremstyle{thmstyleone}%
\theoremstyle{thmstyletwo}%

\theoremstyle{thmstylethree}%

\usepackage{float}   
\usepackage{subfigure} 
\usepackage{indentfirst}

\begin{document}
\title[Simulation of radiation damage effect on silicon detectors using RASER]{Simulation of radiation damage effect on silicon detectors using RASER}


\author[1,2]{\fnm{Xingchen} \sur{Li}}
\author[3,4]{\fnm{Chenxi} \sur{Fu}}
\author[4,5]{\fnm{Hui} \sur{Li}}
\author[4]{\fnm{Zhan} \sur{Li}}
\author[1]{\fnm{Lin} \sur{Zhu}}
\author[4]{\fnm{Congcong} \sur{Wang}}
\author[4]{\fnm{Xiyuan} \sur{Zhang}}
\author[2]{\fnm{Weimin} \sur{Song}}
\author[1]{\fnm{Hui} \sur{Liang}}
\author[1]{\fnm{Cong} \sur{Liu}}
\author[1]{\fnm{Hongbo} \sur{Wang}}


\author*[4]{\fnm{Xin} \sur{Shi}}\email{shixin@ihep.ac.cn}
\author*[1]{\fnm{Suyu} \sur{Xiao}}\email{suyu.xiao@iat.cn}

\affil*[1]{\orgname{Shandong Institute of Advanced Technology}, \orgaddress{\street{1501 Panlong Road}, \city{Jinan}, \postcode{250100}, \state{Shandong}, \country{China}}}

\affil[2]{\orgname{Jilin University}, \orgaddress{\street{NO.2699 Qianjin Street}, \city{Changchun}, \postcode{130216}, \state{Jilin}, \country{China}}}

\affil[3]{\orgname{University of Chinese Academy of Sciences}, \orgaddress{\city{No.19A Yuquan Road},  \city{Shijingshan District}, \postcode{100049}, \state{Beijing}, \country{China}}}

\affil[4]{\orgdiv{Institute of High Energy Physics}, \orgname{Chinese Academy of Science}, \orgaddress{\street{No.19B Yuquan Road}, \city{Shijingshan District}, \postcode{100049}, \state{Beijing}, \country{China}}}

\affil[5]{\orgname{Tsinghua University}, \orgaddress{\city{Haidian  District}, \postcode{100084}, \state{Beijing}, \country{China}}}


\abstract{Silicon detectors play a crucial role in high energy physics experiments. In future high energy physics experiments, silicon detectors will be exposed to extremely high fluence environment, which can significantly affect their performance. It is important to understand the electrical behavior of detectors after irradiation.
In this study, an irradiation simulation framework is constructed in RASER to simulate leakage current and charge collection effciency. The defect parameters are obtained from the Hamburg penta trap model (HPTM). Based on this work, we predict the similar silicon inner tracker which under a ten-year CEPC Higgs mode run can still maintain over $90\%$ charge collection efficiency.
}

\keywords{silicon strip detector, radiation damage,  RASER, HPTM}



\maketitle

\section{Introduction}\label{sec1}
    Silicon detectors are widely used in high energy physics experiments due to their excellent time resolution and spatial resolution. In future hadron colliders\cite{bib1}, silicon detectors will face significant challenges due to large exposure to irradiation. The high fluence irradiation will lead to significant degradation of silicon sensors' performance such as decreasing the charge collection and increasing the leakage current. Over the last twenty years, many studies have observed that irradiated silicon devices exhibit a double peak electrical field \cite{bib2} and severe carrier trapping effect \cite{bib3}. 
    
    In recent years, using Monte Carlo and TCAD simulations to study the performance of detectors after irradiation has become a widely adopted research approach. The Hamburg penta trap model (HPTM) \cite{bib4} defect parameters are used for our irradiation simulation work and adjust the introduction rate using measurement data from ATLAS ITk sensors \cite{bib14}\cite{bib15}.
    
    In the Circular Electron Positron Collider (CEPC) design \cite{bib5}, the pair-production and off-energy beam particles will contribute a $\rm 2\times 10^{13}\ n_{eq}/cm^2$ fluence background on endcap region during Higgs run mode. It is important that the CEPC inner tracker could tolerate this irradiation environment. In this work, we simulate the charge collection efficiency performance of the detector under CEPC irradiation conditions.

\section{Simulation process}\label{sec3}

    RASER \cite{bib6} is an open source software designed to simulate physics processes in semiconductor detector implemented in Python. In previous work, RASER has achieved excellent simulation results for time resolution of SiC PIN\cite{bib7} and 3D\cite{bib8} detectors, and edge-TCT scan of Si LGAD\cite{bib9}. In this work, a simulation framework for irradiated silicon detectors is developed.
    
    First, the open source software DEVSIM\cite{bib10} is used to construct a radiation damage simulation by adding effect from Shockley-Read-Hall (SRH) recombination, effective space charge distribution\cite{bib13} and calculate carrier trapping time. The van Overstraeten impact ionization model\cite{bib11} and the trap-assisted tunneling model\cite{bib12} are added in the leakage current and electric field simulation process.
    
    Then, Geant4\cite{bib17} is used to simulate particles passing through the detectors and depositing energy. After converting energy deposition into number of charge carriers and simulating the carrier drifting process, Shockley-Ramo theorem and carrier trapping mechanism is used to calculate the induced current. 
    
    The defect parameters used in our simulation is from the Hamburg penta trap model (HPTM) \cite{bib4}. Considering that the physics models and many parameters used in our simulation are different from Synopsys TCAD, we modify the introduction rate in the defect parameters based on IV and charge collection data from ATLAS ITk measurement\cite{bib14} \cite{bib15} to better matches experimental results. The parameters are listed in Table \ref{tab2}. 
    
    \begin{table}[h]
    \begin{center}
    \caption{\centering{Defect information of HPTM and this work}}\label{tab2}%
    \begin{tabular}{@{}ccccccc@{}}
    \toprule
    \multirow{2}{*}{Defect} & \multirow{2}{*}{Type}     & \multirow{2}{*}{Energy}   & \multirow{2}{*}{$\sigma_{e} \rm{({cm}^{2}})$} & \multirow{2}{*}{$\sigma_{h} (\rm{{cm}^{2}})$} & \multicolumn{2}{c}{Introduction rate $(\rm{{cm}^{-1}})$} \\
    
    & & & & & origin & this work$^{\star}$ \\
    \midrule
    E30K & Donor    & $E_{c}-0.1$\;eV  & 2.3$\times{10}^{-14}$  & 2.92$\times{10}^{-16}$  & 0.0497 & 2.0810$^{\star}$\\
    $\rm V_{3}$& Acceptor    & $E_{c}-0.458$\;eV   & 2.551$\times{10}^{-14}$  & 1.511$\times{10}^{-13}$ & 0.6447 & 1.6504$^{\star}$ \\
    $\rm I_{p}$& Acceptor    & $E_{c}-0.545$\;eV    & 4.478$\times{10}^{-15}$  & 6.709$\times{10}^{-15}$ & 0.4335 & 0.6936$^{\star}$ \\
    H220&   Donor    & $E_{v}+0.48$\;eV    & 4.166$\times{10}^{-15}$  & 1.965$\times{10}^{-16}$ & 0.5978 & 2.6112$^{\star}$  \\
    $\rm C_{i}O_{i}$& Donor    & $E_{v}+0.36$\;eV   & 3.230$\times{10}^{-17}$  & 2.036$\times{10}^{-14}$ & 0.3780 & 1.6511$^{\star}$  \\
    \botrule
    \end{tabular}
    \end{center}
    \end{table}

\section{Simulation results}\label{sec4}

     In the simulation, the input parameters of the p-bulk region depth is set to 300 µm, and the doping concentration is set to $\rm 4 \times{10}^{12} \ {cm}^{-3}$ \cite{bib16}. The pitch of the strip is set to 75 µm, and the electrode width to 25 µm for the simulation of the weighting field and electric field. The simulation temperature is set to $\rm -20^\circ C$.

\begin{figure}[H]
	\centering  
	\subfigure[]{
		\includegraphics[width=0.45\linewidth]{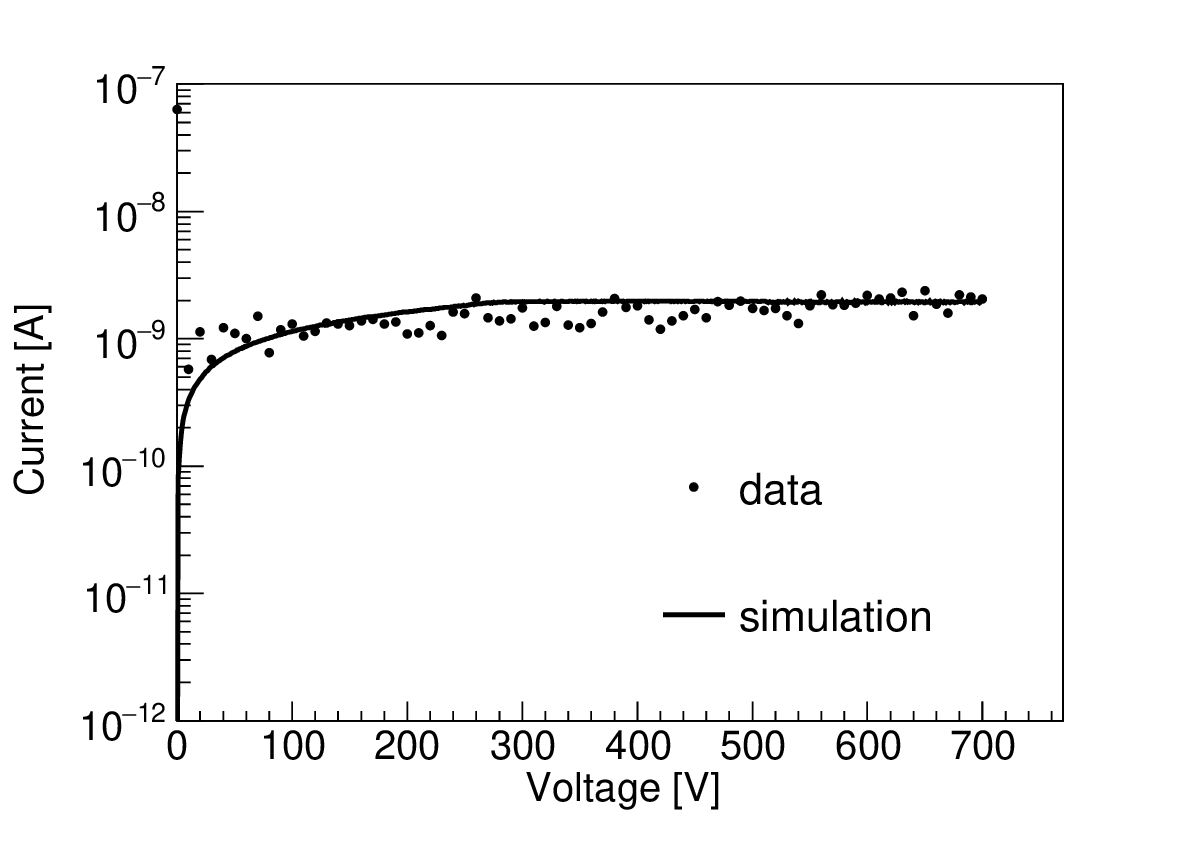}}
	\subfigure[]{
		\includegraphics[width=0.45\linewidth]{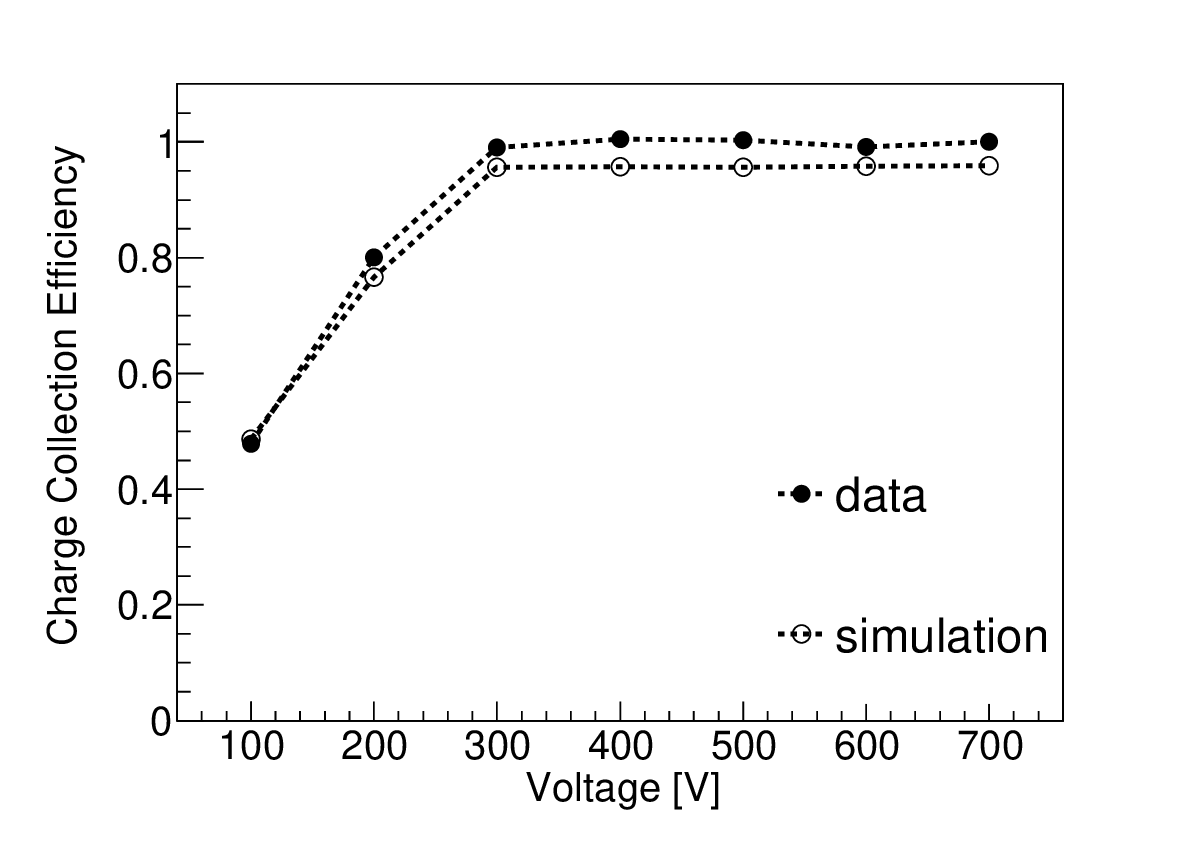}}
	\\
	\caption{\centering{(a) IV results and (b) CCE results comparison between simulations and data without irradiation }}\label{sim_unir}
    \end{figure}

    The data from non-irradiated sensors is used to verify the simulation setup and to calibrate parameters. Figure \ref{sim_unir} shows the simulation results for leakage current and charge collection efficiency which agrees well with data. The charge collection efficiency of $100\%$ is based on data at 700V, which is 23084 e.
    
\begin{figure}[H]
	\centering  
	\subfigure[]{
		\includegraphics[width=0.45\linewidth]{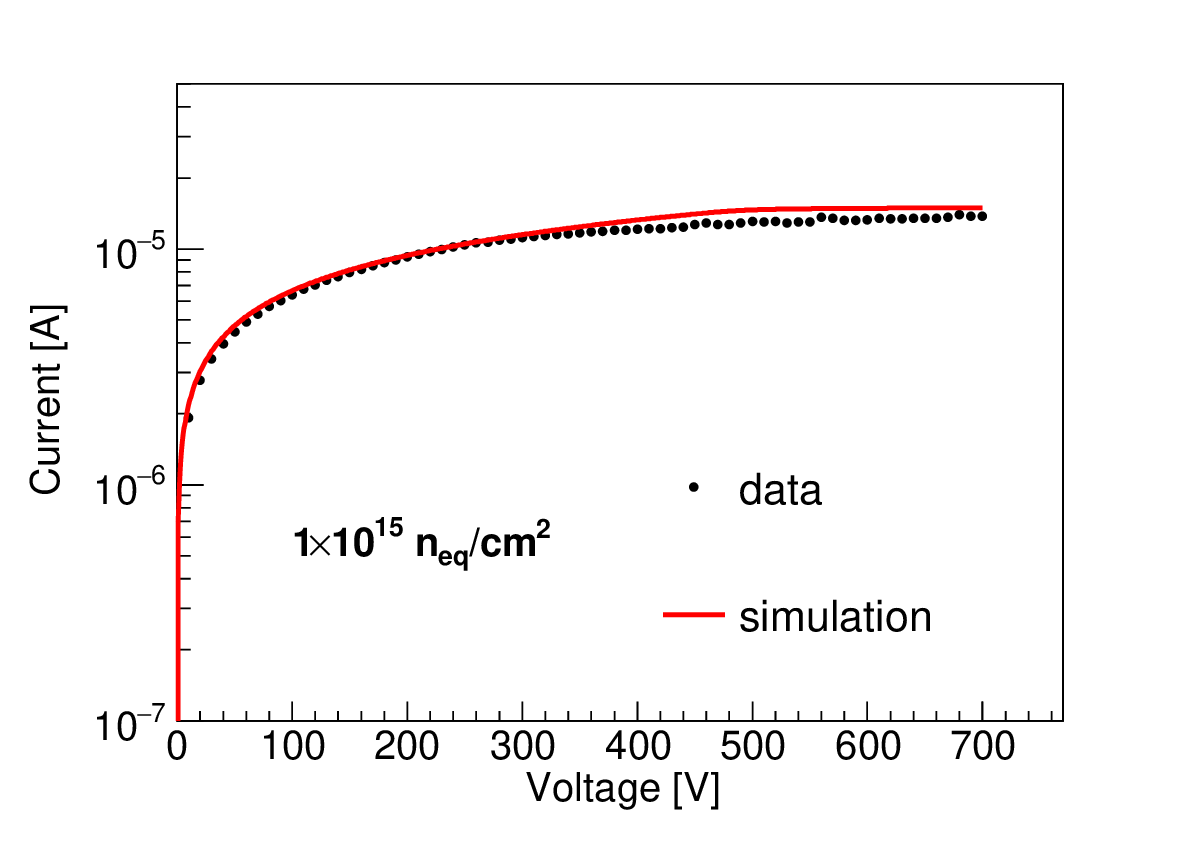}}
	\subfigure[]{
		\includegraphics[width=0.45\linewidth]{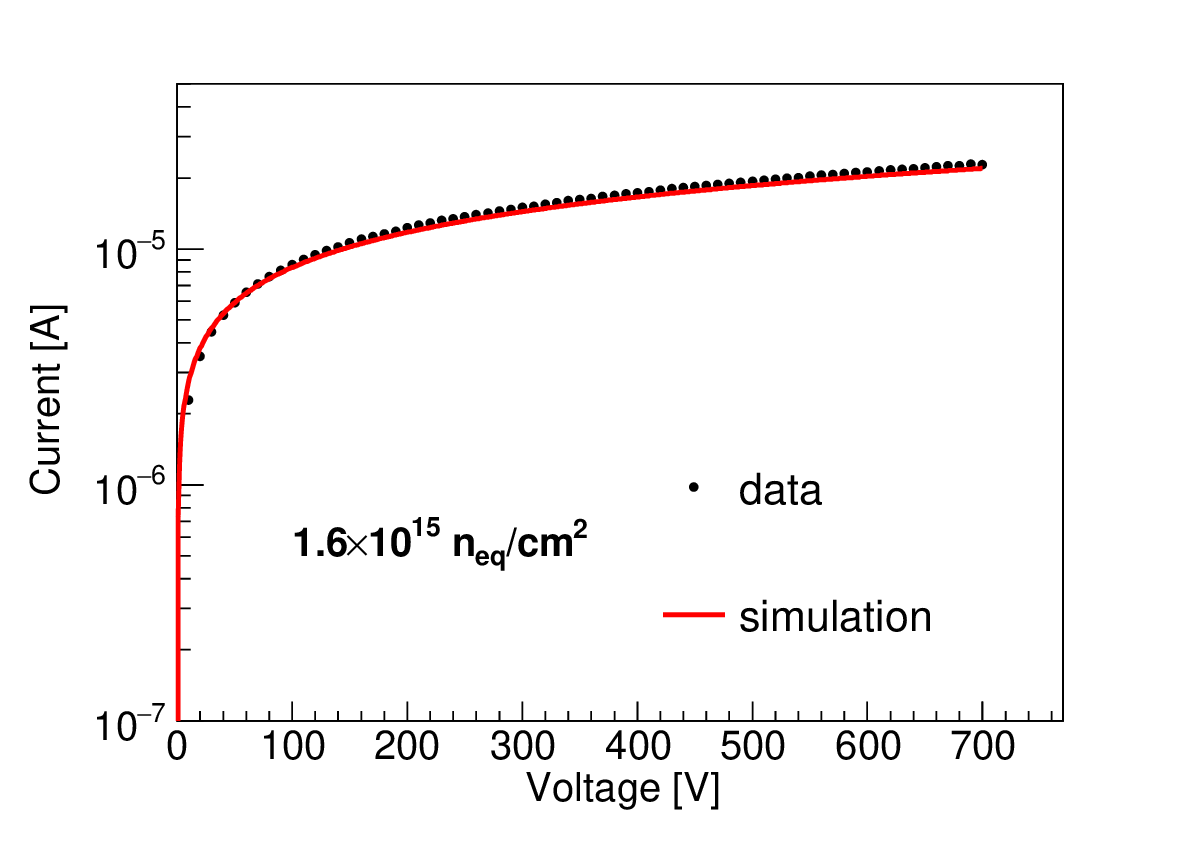}}
	\\
	\caption{\centering{IV results comparisons between data and simulation at (a) $\rm1\times 10^{15}\ n_{eq}/cm^2$ and (b) $\rm1.6\times 10^{15}\ n_{eq}/cm^2$ fluences}}\label{iv}
    \end{figure}

    The defects induced by irradiation significantly increase the recombination rate, which leads to an increase in leakage current. Figure \ref{iv} shows a comparison IV data between the experiment and simulation at fluences of $\rm1\times 10^{15}\ n_{eq}/cm^2$ and $\rm1.6\times 10^{15}\ n_{eq}/cm^2$. The simulation results of leakage current are significantly higher than the non-irradiated case and align closely with the data.

\begin{figure}[H]%
    \centering
    \includegraphics[width=0.7\textwidth]{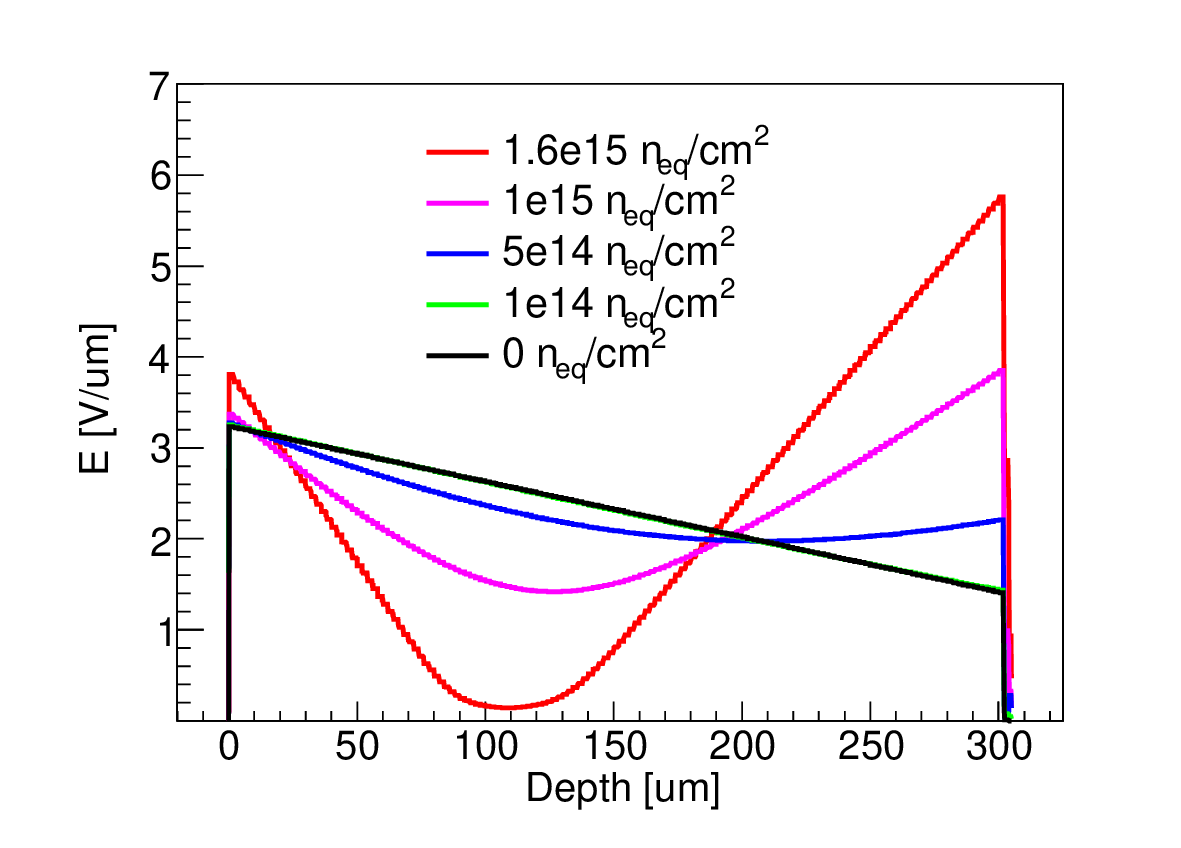}
    \caption{\centering{Simulation results of electrical field distribution with different fluences at 700 V}}\label{e_field_ir}
    \end{figure}

     The electric field is important for comprehending detector's performance and simulating carrier movement. The electric field values at 700 V with different fluences are shown in Figure \ref{e_field_ir}. The double peak electrical field becomes more and more obvious as the fluence increases. \\
    
\begin{figure}[H]%
    \centering
    \includegraphics[width=0.7\textwidth]{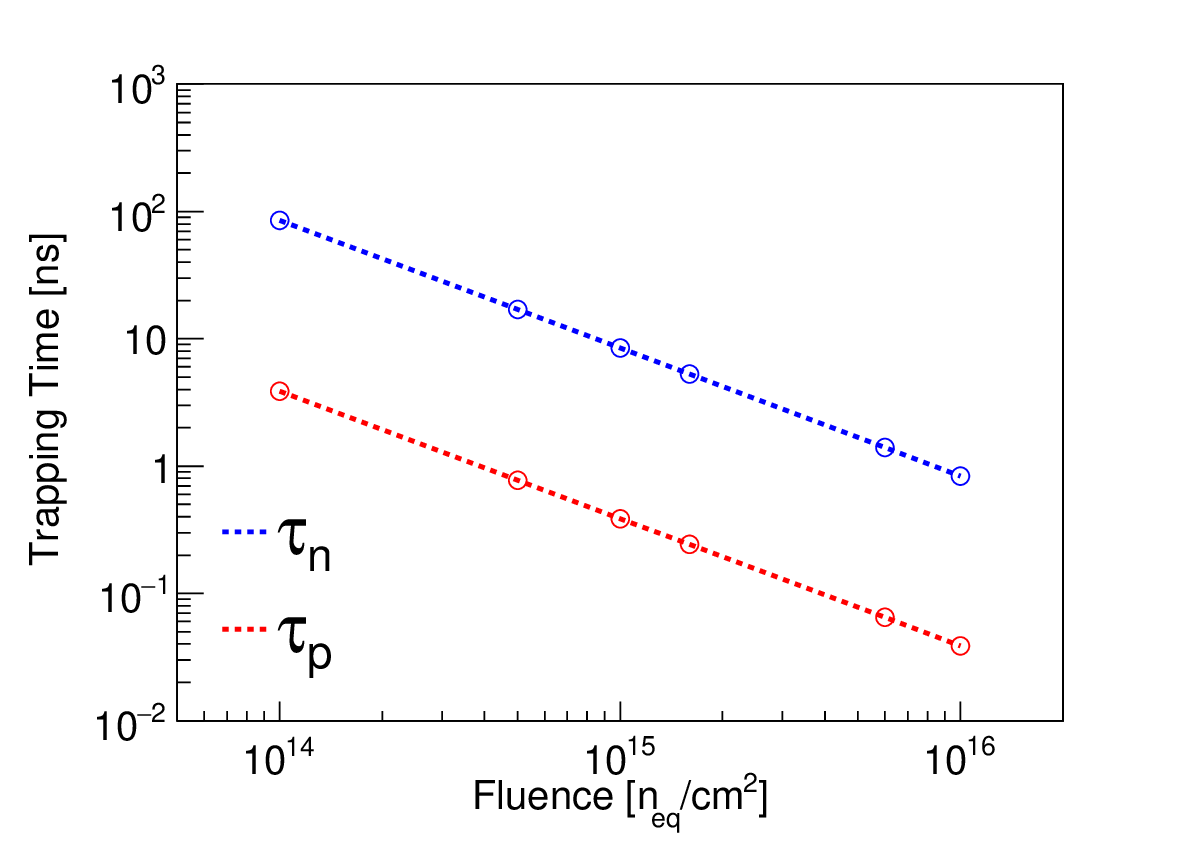}
    \caption{\centering{Simulation results of trapping time with different fluences}}\label{trapping_time_log}
    \end{figure}

     The average trapping time changes with irradiation fluences is illustrated in Figure \ref{trapping_time_log}. The simulation results show a higher capture ability for holes compared to electrons. When the fluence reaches the order of $\rm10^{15}\ n_{eq}/cm^2$, the trapping time approaches the nanosecond level, which can significantly reduce the signal.
    
\begin{figure}[H]
	\centering  
    \includegraphics[width=0.6\textwidth]{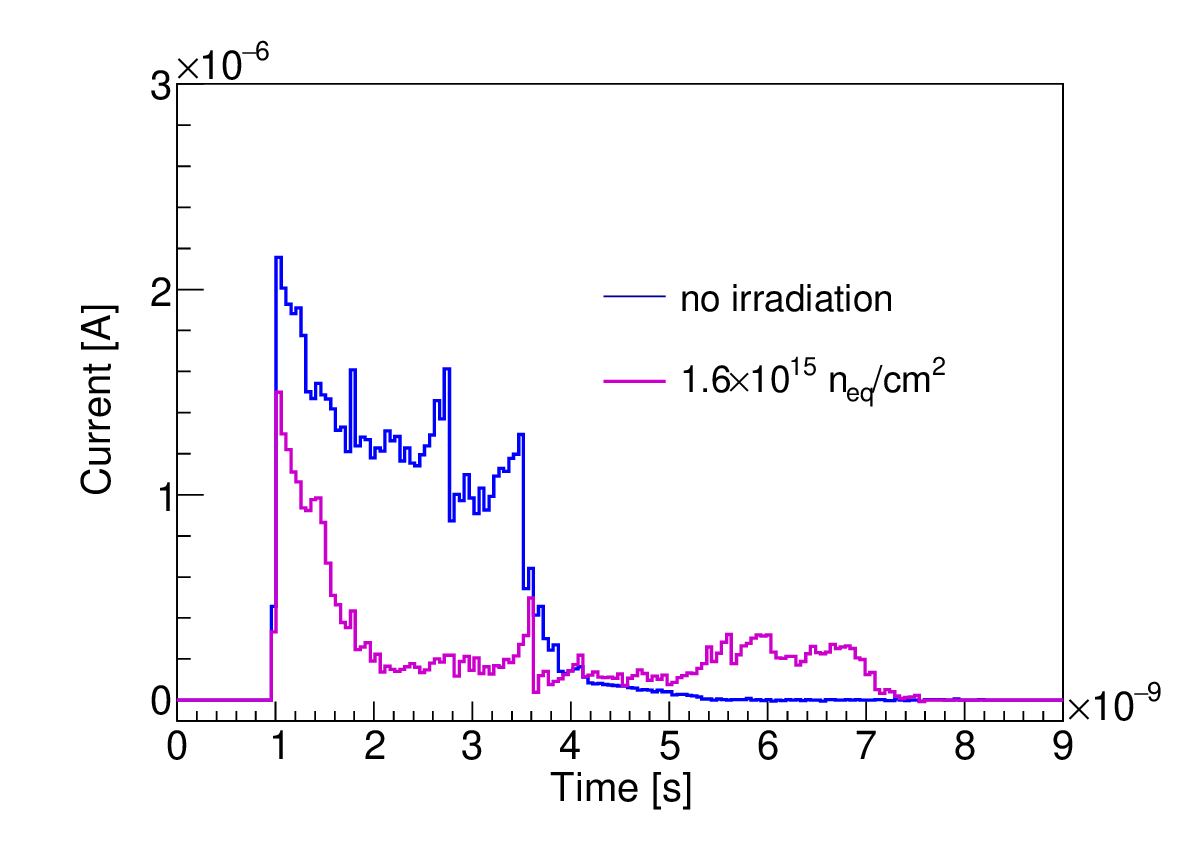}
	\caption{\centering{Simulation of induced current waveform with non-irradiation and $\rm1.6\times 10^{15}\ n_{eq}/cm^2$ fluence at 500 V}}\label{current_ir}
    \end{figure}

     The simulation results of induced current signals produced by the beta source is shown in Figure \ref{current_ir}. The after irradiation signal exhibits a longer fall time and lower amplitude compared to the non irradiation signal, which is associated with trapping effect and low electric field region in the irradiated device.
     
\begin{figure}[H]
	\centering  
	\includegraphics[width=0.6\linewidth]{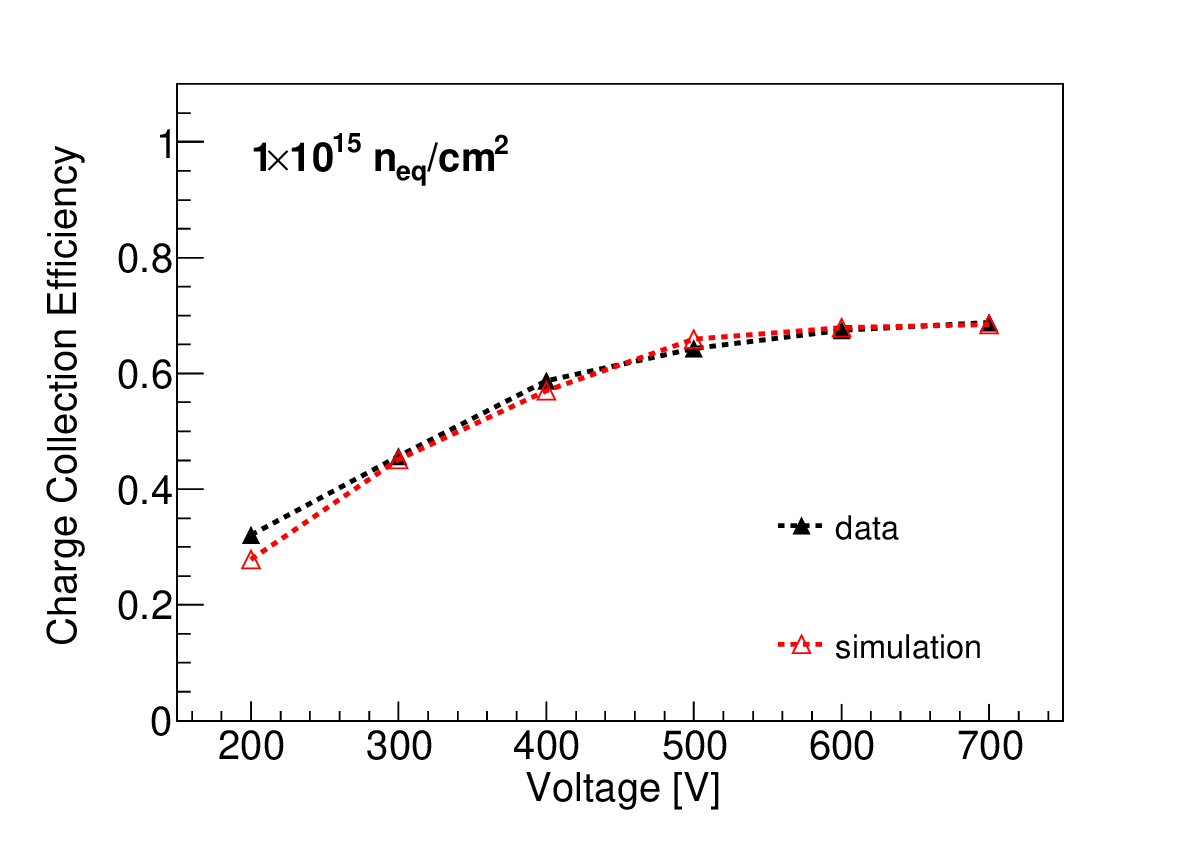}

	\caption{\centering{CCE results comparisons between data and simulation at $\rm1\times 10^{15}\ n_{eq}/cm^2$ fluences}}\label{ir_cce}
    \end{figure}

     The simulation results of charge collection efficiency at different voltages for fluences of $\rm1\times 10^{15}\ n_{eq}/cm^2$ are shown in Figure \ref{ir_cce}. The simulation results demonstrate good agreement with the data at the working voltage.
    
\section{Irradiation prediction in CEPC}\label{sec5}
\begin{figure}[H]
	\centering  
    \includegraphics[width=0.6\textwidth]{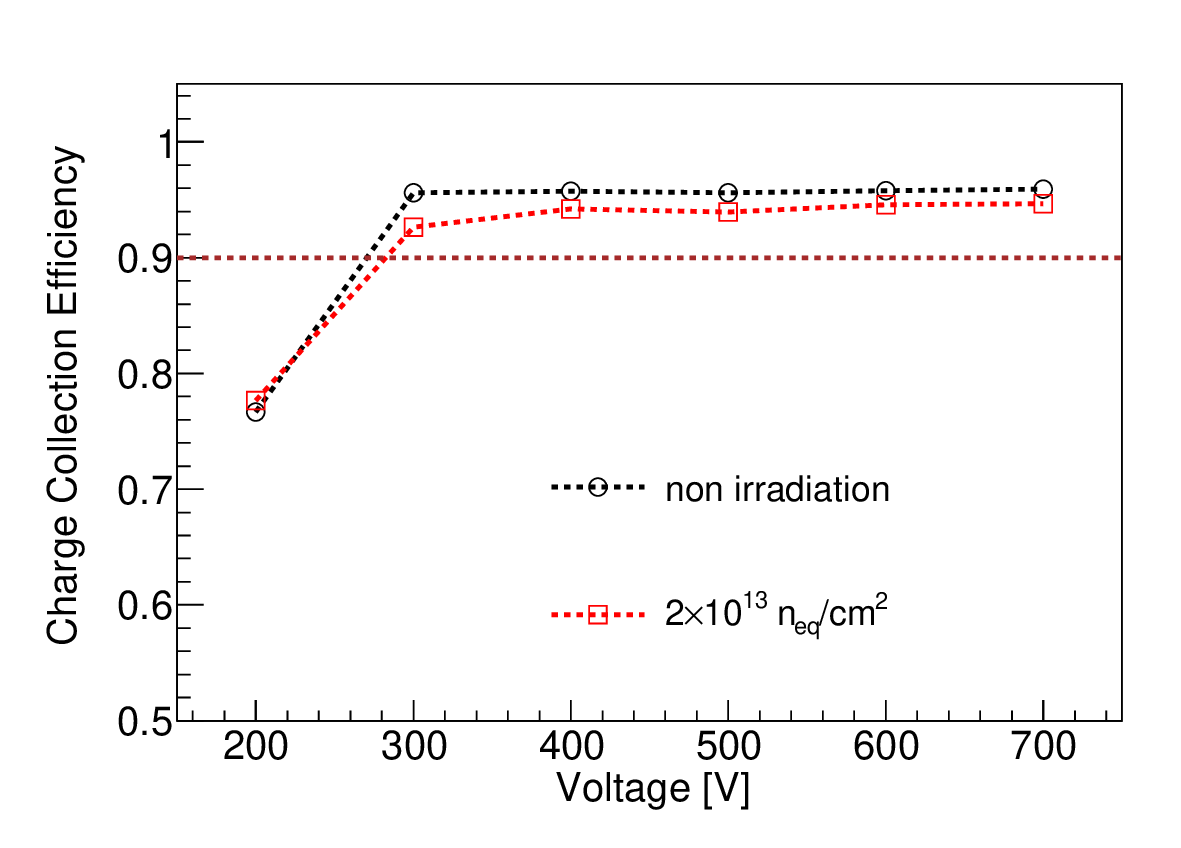}
	\caption{\centering{CCE simulation result under CEPC irradiation background}}\label{1e14}
    \end{figure}

    During a 10-year operation of the CEPC Higgs run, the endcap tracker will face  $\rm 2\times 10^{13}\ n_{eq}/cm^2$ total fluence\cite{bib5}. We simulate the effects of this total irradiation dose. Evaluating the charge collection efficiency of similar silicon sensors at this fluence. Figure \ref{1e14} shows detector could still maintain over $90\%$ charge collection efficiency, meaning that this irradiation background would not significant affect charge collection property.

\section{Conclusion}\label{sec6}
    A simulation of irradiated silicon sensors has been developed using RASER software. The defect parameters of the Hamburg penta trap model have been incorporated, and the defect introduction rate is adjusted by the ATLAS ITk data of leakage current and charge collection efficiency. Based on this work, we predict the CEPC inner tracker endcap performance after a ten-year run in Higgs mode. The simulation results indicate that the detector can maintain over $90\%$ charge collection efficiency.

\bmhead{Acknowledgments}

   The research was supported and financed in large part by the National Key Research and Development Program of China under Grant No. 2023YFA1605902 from the Ministry of Science and Technology, the National Natural Science Foundation of China (W2443002, 12405219, 11961141014 and 12305207), the State Key Laboratory of Particle Detection and Electronics (SKLPDE-KF-202313, SKLPDE-KF-202401 and SKLPDE-ZZ-202312), Natural Science Foundation of Shandong Province Youth Fund (ZR2022QA098) and support from CERN DRD3 Collaboration.
    



\end{document}